\documentstyle[twoside,fleqn,espcrc2]{article}
\input{epsf}
\def\tr{\mathop{\rm tr}\nolimits}
\def\Tr{\mathop{\rm Tr}\nolimits}

\newcommand{\AmS}{{\protect\the\textfont2
  A\kern-.1667em\lower.5ex\hbox{M}\kern-.125emS}}

\title{Applications of the Eigenmodes of the Staggered Dirac Operator}

\author{L. Venkataraman and G. Kilcup\address{Department of Physics,
        The Ohio State University \\
        174  West 18th Ave, Columbus, OH\,\, 43210}}
\begin{document}
\begin{abstract}
We have constructed the lowest few eigenvectors of the staggered Dirac
operator on $SU(3)$ gauge configurations, both quenched and dynamical.
We use these modes to study the topological charge and to construct
approximate hadronic correlators.
\end{abstract}
\maketitle

\section{Introduction}
In a recent study~\cite{lat96} we successfully calculated the mass of
the $\eta'$ meson using staggered fermions.  According to conventional
lore, the $\eta'$ receives a large portion of its mass from instanton
effects. At the fermionic level, this would mean receiving
contributions from zero modes of the Dirac operator.  These zero modes
come with a definite chirality in the continuum and are associated
with the topological charge ($Q$) of a gauge configuration via the
index theorem:
\begin{equation}
Q = n_{+} - n_{-}
\end{equation}
where $n_{+}$ and $n_{-}$ are respectively the number of right and
left-handed zero modes.  At finite lattice spacing, ${\gamma}_5$ and
$\rlap{\,/}D$ do not exactly anti-commute, so there are no exact
chiral zero modes and nor is the topological charge of  gauge
configurations well defined.  Nevertheless, one might expect to find a
few low lying eigenmodes carrying chiral charge.  Smit and
Vink~\cite{remnants} verified this sort of behavior for $U(1)$ gauge
theory in 2 dimensions.

In this work, we set out to study the extent to which the lattice
``zero modes'' reproduce the continuum picture for QCD.  Accordingly,
we have constructed the lowest few eigenmodes of staggered
$\rlap{\,/}D$ on both quenched and dynamical $SU(3)$ gauge
configurations (83 samples each) of size $16^3 \times 32$
corresponding to a lattice spacing of 0.1fm.  We have used the
subspace iterations method~\cite{Bunk} to compute the eigenmodes.  The
eigenvalues that we obtain have relative errors in the range $10^{-7}$
for the smallest through $10^{-4}$ for the highest.

\section{Topological charge}

\begin{figure}[t]
\centerline{\epsfysize=3in\epsfbox{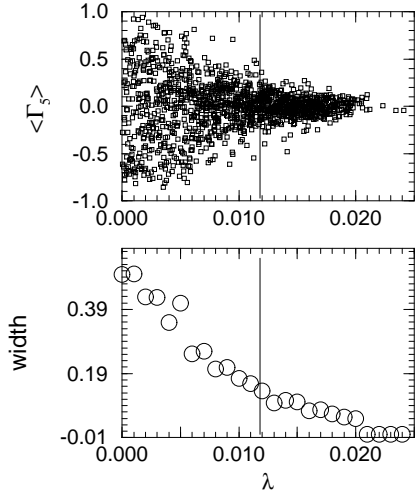}}
\vspace{-1.3cm}
\caption{Upper:$\langle {\Gamma}_5 \rangle$ on the dynamical ensemble. Lower: Fluctuations in $\langle {\Gamma}_5 \rangle$.}
\vspace{-0.52cm}
\label{dg5}
\end{figure}

\begin{figure}[t]
\centerline{\epsfysize=3in\epsfbox{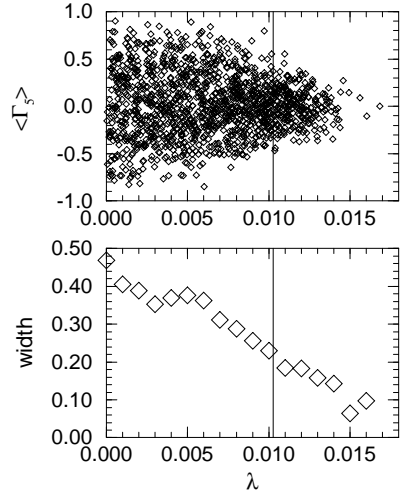}}
\vspace{-1.7cm}
\caption{Above:$\langle {\Gamma}_5 \rangle$ on the quenched ensemble. Below:Fluctuations in $\langle {\Gamma}_5\rangle$. }
\vspace{-.7cm}
\label{qg5}
\end{figure}

The lattice analog of the $U(1)$ axial anomaly equation gives
\begin{equation}
Q  =  m\tr({\Gamma}_5G(x,y))
     \equiv  m\sum_{\lambda}\frac{u^{\dag}_{\lambda}(x){\Gamma}_5\,u_{\lambda}(y)}{i{\lambda}_r + m}
\label{trg5sf1}
\end{equation}
where $G(x,y)$ is the fermion propagator.  The mode by mode
contribution to $u_{\lambda}^{\dag} {\Gamma}_5 u_{\lambda}$ on all the
dynamical configurations is shown in Figure~\ref{dg5}.  It appears to
diminish as $\lambda$ increases which can be verified quantitatively by
calculating the width, shown in the bottom part of figure ~\ref{dg5}.
This behavior is consistent with the continuum where $
u_{\lambda}^{\dag} {\Gamma}_5 u_{\lambda} = 0 $ for all the non zero
eigenvalues.  Figure~\ref{qg5} shows a similar set of plots for the
quenched ensemble.  Although both $\langle {\Gamma}_5 \rangle$ and its
width show qualitatively similar behavior as before, they seem to
drop more slowly with increasing $\lambda$.  The solid line present in
all the figures shown so far represents the minimum value of the 32nd
eigenvalue in the ensemble, ie. the line below which we are sure to have 
found 100\% of the modes.
\begin{figure}[h]
\centerline{\epsfxsize=2.5in\epsfbox{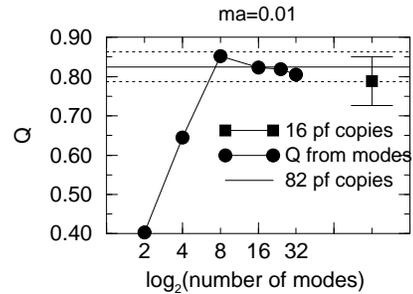}}
\vspace{-1.5cm}
\caption{Topological charge on a typical dynamical configuration}
\label{q05}
\vspace{-0.52cm}
\end{figure}
The sum in eqn(\ref{trg5sf1}) for $r=2,\,4,\,8,\,24$ and 32 modes at
$ma=0.01$ on a typical dynamical configuration is shown in
figure~\ref{q05}.  The plateau in $Q$ as the number of modes increases
implies that the contribution of the higher modes to the sum in
eqn~\ref{trg5sf1} is becoming negligibly small. Although, the plot
shown here is for one configuration, a similar behavior is seen on
all the configurations, quenched and dynamical.  In Figure~\ref{q05},
the isolated point represented by a square is the value of $Q$
obtained from pseudofermions.  The ``true'' value obtained by
averaging over 82 copies of noise lies within the band marked by the
dashed lines.  Since the plateau in $Q$ lies within this band it is a
very good indication that these lowest modes are the topological zero
modes which make the dominant contribution to the sum in
eqn~\ref{trg5sf1}.  We have compared $Q$ obtained from pseudofermions
and modes on all the dynamical configurations and find that both agree
within errors.  On the quenched ensemble, the difference between the
two estimators appears to be statistically significant.  This simply
reinforces the inference reached by comparing Figures~\ref{dg5} and
~\ref{qg5} that more than 32 lowest eigenvalues are required to
identify an upper limit for the number of shifted zero modes on the
quenched configurations.

\section{Hadron Correlators}

Calculation of hadron correlators on the lattice normally involves
inverting $\rlap{\,/}D + m$ to obtain the quark propagator.
When resolved in terms of the eigenvalues and 
eigenvectors of $\rlap{\,/}D$, on a fixed background gauge field,
it takes the form 
\begin{equation}
G(x,y) = \sum_{\lambda} \frac{u_{\lambda}(x) u_{\lambda}^{\dag}(y)}{i\lambda + m}.
\label{Geig}
\end{equation}
Comparing the full propagator obtained  using standard inversion 
procedures like CG, with that constructed out of the sum above
with the available eigenmodes gives an insight into the role 
of the eigenmodes in determining hadronic quantities.   

\begin{figure}[h]
\centerline{\epsfxsize=3.1in\epsfbox{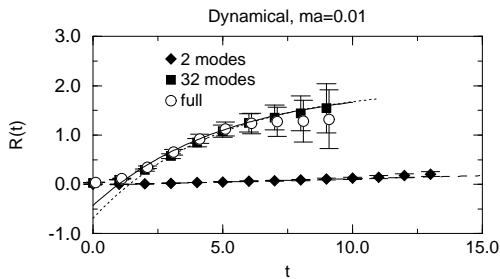}}
\vspace{-1.5cm}
\caption{Pion-$\eta'$ splitting.}
\label{r01}
\vspace*{-0.52cm}
\end{figure}

The $\eta'$ meson, being a flavor singlet, has both connected and
disconnected contributions to its correlator.  The discussion in
section 2 suggests an extension of the analysis to the $\eta'$
disconnected correlator which is given by $\langle
\Tr({\Gamma}_5\,G(x,x))\Tr({\Gamma}_5\,G(y,y))\rangle $.  We calculate
this quantity using pseudofermions and with the available 32 modes.
The ratio $R(t)$ of disconnected to the connected correlator for
$ma=0.01$ is shown in fig~\ref{r01}.  For the connected correlator, we
use the full result available from our previous study~\cite{lat96}.
On the dynamical configurations, $R(t)$ is fit to the
form~\cite{lat96} $R(t)=\frac{N_v}{N_f}[1-\exp(-(m_{\eta'} - m_8)t)]$.
For staggered fermions, $N_v$, the number of valence fermions equals 4
while $m_8^2 \equiv (4m_K^2+3m_{\pi}^2+m_{\eta}^2)/8$.  It is evident
from the figure that $R(t)$ data with 32 modes agree within error with
the full result.
\vspace*{-0.5cm}
We find similarly good agreement with only 24 modes though 16 appears
to be not quite enough.  The pattern is the same at the heavier 
quark masses, 0.02 and 0.03.

We note that this behavior is not automatic as 32 modes are
insufficient to yield a good approximation to the true quark
propagator.  We find, for example, that the pion is off by a large
factor.  Based on our knowledge of the staggered eigenvalue spectrum
and on recent work done by Negele {\it et al.}~\cite{Negele} with
Wilson fermions, one might expect that 100-200 modes would be needed
to reproduce the pion propagator.  Thus the $\eta'$ is special in
exhibiting an extraordinary sensitivity to the lowest few modes.

The quark propagator, constructed out of the available 32 modes,
can be used as a starting seed for calculating $(\rlap{\,/}D+m)^{-1}$
using CG. 
\begin{figure}[t]
\centerline{\epsfxsize=3in \epsfbox{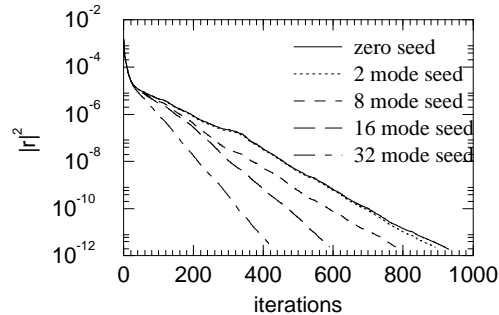}}
\vspace{-1.5cm}
\caption{Comparison of CG convergence.}
\vspace{-0.52cm}
\label{residue}
\end{figure}
The benefit of using such approximate solutions as a starting iterate
in the CG procedure is illustrated in Figure~\ref{residue}. The number
of CG iterations required for the residual to converge to a given
tolerance drops by more than $50\%$ at $ma=0.01$. At higher quark
masses, this effect is not so pronounced owing to the fact that the
lowest 32 eigenvalues are smaller than the quark mass itself.  These
conclusions hold only for delta function sources or gauge fixed wall
sources built out of delta functions.  Quark propagators from noisy
sources are dominated by short distance modes and hence there is
virtually no improvement in the CG convergence rate when the starting
seed is built out of long wavelength modes.

\section{Conclusions}
We find that it is possible to identify the shifted zero modes on the
lattice since they make the dominant contribution to
$u_{\lambda}{\Gamma}_5u_{\lambda}$ and $\tr({\Gamma}_5G(x,y))$.  On
our dynamical configurations, it seems that the number of topological
zero modes must be less than 32 while on the quenched configurations the
data indicates that there are more than this number.  For most
hadronic quantities, the approximate propagator constructed out of
these topological zero modes is not accurate enough to be
useful. However, a remarkable result of our study is that we find that
the $\eta'$ flavor singlet interaction on the dynamical configurations
is dominated entirely by the quasi-zero modes of $\rlap{\,/}D$.

\end{document}